\documentclass[preprint,aps,eqsecnum]{revtex4}

\usepackage{graphicx}
\usepackage{bm}
\usepackage{amsmath}
\usepackage{amssymb}
\usepackage{amsfonts}
\usepackage[dvips]{color}

\newcommand{\bra}{\langle}
\newcommand{\ket}{\rangle}

\newcommand{\dsl}{\partial\kern-0.55em\raise 0.14ex\hbox{/}}

\newcommand{\Vlk}{V_{\mbox{\scriptsize low k}}}
\begin{document}

\preprint{KYUSHU-HET-112}

\title{Problems in the derivations of the renormalization group equation
for the low momentum nucleon interactions}

\author{Koji Harada}
\email{harada@phys.kyushu-u.ac.jp}

\affiliation{Department of Physics, Kyushu University\\
Fukuoka 810-8560 Japan}

\date{\today}

\begin{abstract}
 We carefully examine all the four derivations of the renormalization
 group equation (RGE) for the so-called $\Vlk$ potential, given by
 Bogner, \textit{et.~al.}[nucl-th/0111042]. Two derivations based on the
 ``semi-group composition law'' are shown to be unjustified, while the
 other two based on the completeness relation of the model space must be
 modified if there are bound states. It is however shown that the RGE is
 unchanged if the bound state wavefunctions in the reduced theory are
 required to have the same low-momentum components as those in the
 original theory. Several aspects of the $\Vlk$ approach are also
 discussed.
\end{abstract}

\maketitle

\section{Introduction}

Recently there has been great interest in studying nucleon-nucleon (NN)
interactions by using Effective Field Theory
(EFT)~\cite{Weinberg:1990rz, Weinberg:1991um}. EFT is based on the very
simple idea that the effects of short-distance physics can be simulated
by local (contact) interactions, so that a typical (nuclear) EFT-based
potential consists of a series of local interactions (delta function
potentials and their derivatives) and a long-distance part (Yukawa
potential due to one-pion exchange). EFT is valid only below a certain
scale, the cutoff, and the coupling constants of the local interactions
are determined as functions of the cutoff so that the physical
observables are independent of the cutoff. Lowering the cutoff amounts
to the ``integrating'' the short-distance fluctuations, leading to the
change of the coupling constants of the local interactions. This is
nothing but Wilsonian renormalization group (RG)
idea~\cite{Wilson:1973jj}. EFT together with the RG idea is a promising
alternative to the conventional approach based on phenomenological NN
potentials. It is a model independent, systematically improvable
approach, and is related to QCD through chiral symmetry. See
Refs.~\cite{Beane:2000fx, Bedaque:2002mn, Epelbaum:2005pn,
Machleidt:2007ms} for reviews.

One might however think that the use of the ``realistic''
potentials~\cite{Stoks:1994wp, Wiringa:1994wb, Machleidt:1995km,
Machleidt:2000ge}, which describe thousands of data with $E_{lab}\alt
350$ MeV very accurately, together with the RG idea, would be
\textit{more} effective, though such an approach is inevitably
phenomenological. The so-called $\Vlk$ approach~\cite{Bogner:2000js,
Bogner:2001gq, Bogner:2001jn, Bogner:2002yw, Bogner:2003wn,
Nogga:2004ab} (in its original form) exactly goes along this strategy.

The $\Vlk$ approach originally emerged as an application of the model
space reduction methods developed in shell-model
theory~\cite{Bogner:2000js}. The model space reduction methods
apparently resemble RG transformations; the effects of the higher-energy
states are taken into account as effective interactions. It has been
shown that ``realistic'' potentials evolve to a universal low-energy
potential by the model space reduction~\cite{Bogner:2001gq}. It appears
to conform to the general idea of ``universality'' in the RG theory. On
the other hand, the universal potential may be useful also in a
practical sense, because it is much smoother than the ``realistic''
potentials so that one may avoid calculating the Brueckner
G-matrix~\cite{PhysRev.100.36, PhysRev.103.1353, ProcRSoc.A238.551,
ProcRSoc.A239.267}, which is energy and nucleus dependent.

A relatively simple RG equation (RGE), which is claimed to be satisfied
by the $\Vlk$, has been known;
\begin{equation}
 \frac{d}{d\Lambda}\Vlk^\Lambda(k',k)
  =\frac{2}{\pi}\frac{\Vlk^\Lambda(k',\Lambda)T^\Lambda(\Lambda, k; \Lambda^2)}
  {1-(k/\Lambda)^2},
  \label{eq:vlowk}
\end{equation}
where $\Lambda$ is the (floating) cutoff, $k$ and $k'$ are the relative
momentum of incoming and outgoing nucleons respectively, and $T$ is the
T-matrix.  The superscript $\Lambda$ indicates that the quantity is
cutoff dependent. Note that we put the superscript $\Lambda$ on the
left-on-shell T-matrix to indicate explicitly that it is, unlike the
right-on-shell T-matrix, a $\Lambda$-dependent quantity. It is important
to note that ``the RG equation lies at the heart of the
approach~\cite{Bogner:2001gq}.'' There are several model space
techniques and, according to Refs.~\cite{Bogner:2000js, Bogner:2001jn,
Bogner:2002yw, Bogner:2003wn, PhysRevC.54.684}, they all satisfy the
same RG equation. In this sense, it is considered that they are all
equivalent and the universal low-energy potential is unique. Because
this is a very nontrivial assertion, it is important to examine the
derivations of the RG equation carefully.

To our best knowledge, however, the derivations of the RG equations
presented in Ref.~\cite{Bogner:2001jn} do not seem to have been
scrutinized, though there are many applications. In this paper, we show
that each of the four derivations presented in Ref.~\cite{Bogner:2001jn}
contains a serious defect. Our arguments are not technically intricate
nor too mathematical so that those who are familiar with the derivations
can easily follow them.

We do not claim that their RGE (\ref{eq:vlowk}) is wrong; we only claim
that all the existent \textit{derivations} are either unjustified or, if
there are bound states, insufficient.  It is however shown that extra
terms arising from the existence of the bound states cancel, keeping
Eq.~(\ref{eq:vlowk}) unmodified~\cite{BFS}, if the bound state
wavefunctions in the reduced theory are required to have the same
low-momentum components as those in the original theory.

We notice that most of the numerical results given in the literature are
not derived directly from the RGE, but from one of the model-space
reduction methods such as Lee-Suzuki (LS) method~\cite{Suzuki:1980yp} or
Kuo-Lee-Ratcliff (KLR) method~\cite{NuclPhysA.176.65}. We emphasize that
numerical results obtained with these methods are not affected by the
present work. The procedure is independent of the actual form of the
RGE, and the effective interactions obtained by these methods in fact
satisfy the condition mentioned above, and hence Eq.(\ref{eq:vlowk}).

We assume that the reader is familiar with Ref.~\cite{Bogner:2001jn},
though we recapitulate the main points. We follow their notations as
closely as possible. In several places, we need to supplement their
arguments by providing the missing intermediate steps which we think
they actually took.

We start with the definition of the so-called $\hat{Q}$-box and the KLR
method in Sec.~\ref{sec:KLR}. By supplementing the intermediate steps,
we point out explicitly where they took a wrong step, invalidating their
second and third derivations. In Sec.~\ref{sec:completeness}, we show
that their completeness relation in the ``model space'' is wrong when
bound states are present, invalidating the first derivation. It is
however shown that, if the bound state wavefunctions in the reduced
theory have the same low-momentum components as those in the original
theory, the extra terms cancel each other, and the RGE is not affected
by the presence of bound states. The fourth derivation is also based on
the same completeness relation, so that it is also invalid in the
presence of bound states as we show in Sec.~\ref{sec:spectral}. Again,
the final result, Eq.~(\ref{eq:vlowk}), can be derived without change
under the same condition as in Sec.~\ref{sec:completeness}. In
Sec.~\ref{sec:summary}, we summarize our results and discuss several
related issues.

\section{$\hat{Q}$-box and the Kuo-Lee-Ratcliff method}
\label{sec:KLR}

Let us start with the ``bare'' Lippmann-Schwinger equation for the
(off-shell) T-matrix,
\begin{equation}
 T=V_{NN}+V_{NN}G_0T,
\end{equation}
where $V_{NN}$ is a ``realistic'' potential. In the partial-wave
notation, this may be written as
\begin{equation}
 T(k',k; \omega) = V_{NN}(k',k)
  +\frac{2}{\pi}\mathcal{P}\int_0^\infty 
  V_{NN}(k,p)\frac{p^2 dp}{\omega-p^2}T(p,k; \omega),
\end{equation}
where, following Ref.~\cite{Bogner:2001jn}, we consider the standing
wave boundary condition, but the following arguments are largely
independent of it. (In this paper, we use units with $\hbar=c=M_N=1$.)
Physically speaking, the upper limit should be considered as a very
large cutoff $\Lambda_0$ rather than the infinity.

One would like to replace the ``bare'' Lippmann-Schwinger equation by a
cutoff one, which describes the same physics at low energies. For this
purpose, let us introduce the projection operators, $P$ and $Q$,
satisfying $P+Q=1$, $PQ=QP=0$, $P^2=P$, and $Q^2=Q$. The operator $P$,
which projects states to the low-momentum space, may be written as
\begin{equation}
 P=\frac{2}{\pi}\int_0^\Lambda |k\ket k^2 dk \bra k|,
\end{equation}
where $|k\ket$ stands for the plane wave in the partial-wave notation
with the normalization,
\begin{equation}
 \bra k' | k \ket=\frac{\pi}{2}\frac{\delta(k'-k)}{k'k}.
\end{equation}

The so-called $\hat{Q}$-box may be defined as
\begin{equation}
 \hat{Q}^\Lambda(k',k;\omega)=V_{NN}(k',k)
  +\frac{2}{\pi}\mathcal{P}\int_\Lambda^\infty 
  V_{NN}(k,p)\frac{p^2 dp}{\omega-p^2}\hat{Q}^\Lambda(p,k; \omega),
  \label{Qbox-VNN}
\end{equation}
which is the part of the T-matrix, all of the intermediate states of
which are in the $Q$-space. We put the superscript $\Lambda$ to indicate
that it depends on the cutoff. By using $\hat{Q}$-box, the T-matrix may
be written as
\begin{equation}
 T(k',k;\omega)=\hat{Q}^\Lambda(k',k;\omega)
  +\frac{2}{\pi}\mathcal{P}\int_0^\Lambda 
  \hat{Q}^\Lambda(k',p;\omega)\frac{p^2 dp}{\omega-p^2}T(p,k; \omega).
  \label{T-Qbox}
\end{equation}
Note that $\hat{Q}$-box is energy dependent. If we restrict $k',\ k \le
\Lambda$, then $\hat{Q}$-box may be identified with the Wilsonian
effective potential $V^\Lambda_{WRG}$ satisfying 
\begin{equation}
 \frac{d}{d\Lambda}V^\Lambda_{WRG}(k',k;\omega)=\frac{2}{\pi}
  \frac{V^\Lambda_{WRG}(k',\Lambda;\omega)V^\Lambda_{WRG}(\Lambda,k;\omega)}
  {1-\omega/\Lambda^2},
  \label{wilsonianRGE}
\end{equation}
considered by Birse \textit{et. al.}~\cite{Birse:1998dk}. See 
Refs.~\cite{Harada:2005tw, Harada:2006cw, Harada:2007ua} for the field
theoretical formulation. See also Ref.~\cite{Nakamura:2005uw} for the
comparison of $V_{WRG}$ with $\Vlk$.

First of all, it is important to note that for the \textit{fully
off-shell} T-matrix there is \textit{no} such energy-independent
potential $V^\Lambda$ that satisfies
\begin{equation}
 T(k',k;\omega)=V^\Lambda(k',k)
  +\frac{2}{\pi}\mathcal{P}\int_0^\Lambda 
  V^\Lambda(k',p)\frac{p^2 dp}{\omega-p^2}T(p,k; \omega).
  \label{nosuchV}
\end{equation}
Actually, the $\Lambda$-independence of the (off-shell) T-matrix leads
to the RGE (\ref{wilsonianRGE}), which cannot be satisfied by an
energy-independent potential.

On the other hand, for the \textit{half-on-shell} T-matrix, there exists
an energy-independent potential $\Vlk$,
\begin{equation}
 T(k',k;k^2)=\Vlk^\Lambda(k',k)
  +\frac{2}{\pi}\mathcal{P}\int_0^\Lambda 
  \Vlk^\Lambda(k',p)\frac{p^2 dp}{k^2-p^2}T(p,k; k^2).
  \label{T-Vlk}
\end{equation}
This may serve as the definition of $\Vlk$.

A concrete procedure of constructing such a potential is given by the
KLR folded diagram theory~\cite{NuclPhysA.176.65}. Starting with
Eq.~(\ref{T-Qbox}) with $\omega=k^2$, we have
\begin{equation}
 T(k',k;k^2)=\hat{Q}^\Lambda(k',k;k^2)
  +\frac{2}{\pi}\mathcal{P}\int_0^\Lambda 
  \hat{Q}^\Lambda(k',p;k^2)\frac{p^2 dp}{k^2-p^2}T(p,k; k^2),
\end{equation}
which should be compared with Eq.~(\ref{T-Vlk}). Although
$\hat{Q}^\Lambda(k',k; k^2)$ almost satisfies Eq.~(\ref{T-Vlk}) for
$\Vlk^\Lambda$, $\hat{Q}^\Lambda(k',p;k^2)$ in the integrand depends on
three variables. We therefore write
\begin{equation}
 \Vlk^\Lambda(k',k)=\hat{Q}^\Lambda(k',k; k^2) 
  + \sum_{i=2}^\infty \Delta V^{(i)}(k',k),
\end{equation}
where $\Delta V^{(i)}$ is the correction term coming from the folded
diagrams and is of order of the $i$-th power of $\hat{Q}$.

Substituting it into Eq.~(\ref{T-Vlk}) and comparing with
Eq.~(\ref{T-Qbox}), we have
\begin{align}
 \sum_{i=2}^\infty \Delta V^{(i)}(k',k)
  \!=\!
 &\int_p \frac{\Delta
  Q^\Lambda_{(k)}(k',p)}{k^2-p^2}\hat{Q}^\Lambda(p,k)
  +\int_p\int_{p'}
  \frac{\Delta Q^\Lambda_{(k)}(k',p')\Delta Q^\Lambda_{(k)}(p',p)}
  {(k^2-p^2)(k^2-{p'}^2)}T(p,k;k^2)
 \nonumber \\
 &+\!\int_p\int_{p'}
 \frac{\Delta Q^\Lambda_{(k)}(k',p')\hat{Q}^\Lambda(p',p;p^2)}
  {(k^2-p^2)(k^2-{p'}^2)}T(p,k;k^2)
 \!-\!\!
 \int_p\sum_{i=2}^\infty \Delta V^{(i)}(k',p)\frac{T(p,k;k^2)}{k^2-p^2},
 \label{DeltaV-iterative}
\end{align}
where the notations $\int_p\equiv \frac{2}{\pi} \mathcal{P}
\int_0^\Lambda p^2dp$ and
\begin{equation}
 \Delta Q^\Lambda_{(k)}(k',p) \equiv 
  \hat{Q}^\Lambda(k',p; k^2) - \hat{Q}^\Lambda(k',p; p^2)
\end{equation}
are introduced.  Eq.~(\ref{DeltaV-iterative}) determines $\Delta
V^{(i)}$ iteratively.

Having shown the existence of the $\Vlk^\Lambda$ which satisfies
Eq.~(\ref{T-Vlk}), we now turn to the question about what RGE it
satisfies. The authors of Ref.~\cite{Bogner:2001jn} emphasized the
``semi-group composition law,'' and used the following relation,
\begin{equation}
 \hat{Q}^{\Lambda-\delta \Lambda}(k',k; p^2) = \Vlk^\Lambda(k',k)
  -\delta \Lambda \frac{2}{\pi}\frac{\Vlk^\Lambda(k',
  \Lambda)\Vlk^\Lambda(\Lambda, k)}{1-(p/\Lambda)^2} 
  + \mathcal{O}(\delta \Lambda^2).
  \label{wrong-Qbox}
\end{equation}
This infinitesimal form is derived from the following finite form,
\begin{equation}
 \hat{Q}^{\bar{\Lambda}}(k',k,p^2)=\Vlk^{\Lambda}(k',k)
  +\frac{2}{\pi} \mathcal{P}\int^{\Lambda}_{\bar{\Lambda}}
  \Vlk^{\Lambda}(k',\bar{k})\frac{\bar{k}^2d\bar{k}}{p^2-\bar{k}^2}
  \hat{Q}^{\bar{\Lambda}}(\bar{k},k,p^2),
  \label{twoLambdas}
\end{equation}
where $\bar{\Lambda} \le \Lambda$ is another scale. One might think that
this is the result of integrating Eq.~(\ref{Qbox-VNN}) up to a certain
scale and replacing the ``bare'' potential $V_{NN}(k',k)$ with the
``effective'' one, $\Vlk^\Lambda(k',k)$,
accordingly. Eq.~(\ref{wrong-Qbox}) 
can be easily derived by putting $\bar{\Lambda}=\Lambda
-\delta \Lambda$.  

It is however easy to see that \textit{there is no such}
$\Vlk^{\Lambda}$ \textit{that satisfies} Eq.~(\ref{twoLambdas})
\textit{for an arbitrary ${\Lambda}< \Lambda_0$.}  Consider the limit
$\bar{\Lambda} \rightarrow 0$. From Eq.~(\ref{Qbox-VNN}) we see
\begin{equation}
 \lim_{\bar{\Lambda}\rightarrow 0} \hat{Q}^{\bar{\Lambda}}(k',k; p^2) 
  = T(k', k; p^2),
\end{equation}
so that Eq.~(\ref{twoLambdas}) becomes Eq.~(\ref{nosuchV}) with
$V^\Lambda(k',k)$ being replaced by $\Vlk(k',k)$, but, as we emphasized,
there is no such energy-independent potential that satisfies
Eq.~(\ref{nosuchV}). We have thus shown that Eq.~(\ref{wrong-Qbox})
cannot be valid. This is the defect in the second derivation in
Ref.~\cite{Bogner:2001jn}. The same Eq.~(\ref{wrong-Qbox}) is used in
the third derivation too in a crucial way, so that the third one has the
same defect.

\section{Completeness in the model space}
\label{sec:completeness}

Since $\Vlk$ preserves the half-on-shell T-matrix, one may think of
another way of deriving the RGE, namely, from the $\Lambda$-independence
of the half-on-shell T-matrix in Eq.~(\ref{T-Vlk}). In the case of
$V_{WRG}$, the RGE can be easily derived from the $\Lambda$-independence
of the fully off-shell T-matrix. On the other hand, one cannot follow
the similar manipulation for $\Vlk$, and the authors of
Ref.~\cite{Bogner:2001jn} invoked information on the cutoff state
vectors. But, as we will show, the relation they used is wrong when 
bound states are present. 

Consider the ``bare'' Schr\"odinger equation,
\begin{equation}
(H_0+V_{NN}) |\Psi_k\ket= k^2 |\Psi_k \ket,
 \label{Schroedinger}
\end{equation}
for a scattering state with energy $0 \le k^2< \Lambda^2$, and the
corresponding reduced one,
\begin{equation}
 (H_0 + \Vlk^\Lambda) |\chi^\Lambda_k \ket 
  = k^2 | \chi^\Lambda_k \ket. 
  \label{cutoffSchroedinger}
\end{equation}

The potential $\Vlk^\Lambda$ is a $P$-to-$P$ operator, which may be
constructed by the KLR folded diagram theory or the LS similarity
transformation. In the both cases, the state $|\chi^\Lambda_k \ket$ is
related to the original state by the projection,
\begin{equation}
 |\chi^\Lambda_k \ket = P |\Psi_k \ket.
  \label{stateprojection}
\end{equation}
This condition is nothing but the $\Lambda$-independence of the
half-on-shell T-matrix in the reduced theory. Actually, by applying
$\bra k'| $ with $k' < \Lambda$ to Eqs.~(\ref{Schroedinger}) and
(\ref{cutoffSchroedinger}), we get
\begin{equation}
 \bra k'|V_{NN} |\Psi_k\ket =(k^2-{k'}^2)\bra k'|\Psi_k\ket
  \quad\mbox{and}\quad
  \bra k'|\Vlk |\chi^\Lambda_k\ket =(k^2-{k'}^2)\bra k'|\chi^\Lambda_k\ket,
\end{equation}
but the left hand sides of these equations are by assumption equal to
the half-on-shell T-matrix $T(k',k; k^2)$. Therefore the right hand
sides are also equal, and taking into account that $k'$ is arbitrary, we
see that Eq.~(\ref{stateprojection}) is valid. On the other hand, the
preservation of the half-on-shell T-matrix follows
Eq.~(\ref{stateprojection}).

Their first derivation goes as follows: From the cutoff Schr\"odinger
equation (\ref{cutoffSchroedinger}), one gets the corresponding
Lippmann-Schwinger equation,
\begin{equation}
 |\chi^\Lambda_k\ket = |k\ket + \frac{2}{\pi}\mathcal{P}\int_0^\Lambda
  dp|p\ket\frac{p^2}{k^2-p^2}T(p, k; k^2),
\end{equation}
where $k$ is assumed to be small, $k<\Lambda$. Note that the upper limit
of the integral is $\Lambda$ to be consistent with
Eq.(\ref{stateprojection}). Requiring that the half-on-shell T-matrix is
$\Lambda$-independent,
\begin{equation}
 \frac{d}{d\Lambda}T(k',k;k^2)=
  \frac{d}{d\Lambda}\bra k'|\Vlk^\Lambda|\chi^\Lambda_k\ket=0,
  \label{hosTinv}
\end{equation}
for $k'< \Lambda$ and using
\begin{equation}
 \frac{d}{d\Lambda}|\chi^\Lambda_k\ket 
  = \frac{2}{\pi}|\Lambda \ket
  \frac{\Lambda^2}{k^2-\Lambda^2}T(\Lambda, k; k^2),
  \label{derstate}
\end{equation}
one gets
\begin{equation}
 0=\bra k'|\frac{d\Vlk^\Lambda}{d\Lambda}|\chi^\Lambda_k\ket
  +\frac{2}{\pi}\bra k'|\Vlk^\Lambda|\Lambda \ket
  \frac{\Lambda^2}{k^2-\Lambda^2}T(\Lambda, k; k^2).
  \label{deriv1}
\end{equation}
  The last term may be rewritten by using
\begin{equation}
 \frac{\Lambda^2}{k^2-\Lambda^2}T(\Lambda, k; k^2)
  =\bra \Lambda|\Vlk^\Lambda \frac{\Lambda^2}{H_0+\Vlk^\Lambda
  -\Lambda^2}
  |\chi^\Lambda_k\ket.
\end{equation}

Let us introduce the operator,
\begin{equation}
 J\equiv \frac{2}{\pi} \int_0^\Lambda | \chi^\Lambda_k \ket k^2 dk
  \bra \tilde{\chi}^\Lambda_{k}|,
  \label{J}
\end{equation}
where we introduced a bi-orthogonal basis $\bra \tilde{\chi}_k^\Lambda|$
for each state $|\chi_k^\Lambda\ket$,
\begin{equation}
 \bra \tilde{\chi}^\Lambda_{k'}| \chi^\Lambda_k \ket 
  = \frac{\pi}{2}\frac{\delta(k'-k)}{k'k},
\end{equation}
because $\Vlk^\Lambda$ is not Hermitian.
Note that $J$ is a projection operator, $J^2=J$. One thus obtains 
\begin{equation}
 \bra k' |\frac{d\Vlk^\Lambda}{d\Lambda}J |k\ket
  =-\frac{2}{\pi}\bra k' | \Vlk^\Lambda |\Lambda\ket
  \bra \Lambda |\Vlk^\Lambda \frac{\Lambda^2}{H_0+\Vlk^\Lambda
  -\Lambda^2} J|k\ket.
\end{equation}
If $J$ were the identity operator in the $P$-space, as the authors of
Ref.~\cite{Bogner:2001jn} claimed, then the identity ,
\begin{equation}
 \bra \Lambda | \Vlk^\Lambda \frac{\Lambda^2}{H_0+\Vlk^\Lambda
  -\Lambda^2} |k\ket 
  = T^\Lambda(\Lambda, k; \Lambda^2)\frac{\Lambda^2}{k^2-\Lambda^2 },
\end{equation}
would lead us to
\begin{equation}
 \frac{d}{d\Lambda}\Vlk^\Lambda(k',k)=-\frac{2}{\pi}
  \Vlk^\Lambda(k',\Lambda)T^\Lambda(\Lambda, k; \Lambda^2)
  \frac{\Lambda^2}{k^2-\Lambda^2 },
\end{equation}
which is nothing but Eq.~(\ref{eq:vlowk}). 

In the presence of bound states, however, the operator $J$ cannot be the
identity operator, 
\begin{equation}
 J+
  \sum_i
  \left|\chi^\Lambda_{B_i}\ket
   \bra \tilde{\chi}^\Lambda_{B_i}\right| 
  = P,
  \label{correctcomplete}
\end{equation}
where $|\chi^\Lambda_{B_i}\rangle$ are state vectors for the bound
states in the reduced theory,
\begin{equation}
 (H_0 + \Vlk^\Lambda) |\chi^\Lambda_{B_i} \ket 
  = -k_{B_i}^2 | \chi^\Lambda_{B_i} \ket,
  \label{eq:BS}
\end{equation}
and $\bra \tilde{\chi}_{B_i}^\Lambda|$ is its conjugate bi-orthogonal
state,
\begin{equation}
 \bra \tilde{\chi}_{B_i}^\Lambda|\chi_{B_j}^\Lambda\ket=\delta_{ij}.
  \label{biorthonormal}
\end{equation}

Since the first derivation in Ref.~\cite{Bogner:2001jn} uses the claim
that $J$ is the identity operator in the $P$-space in an essential way,
it contains a serious defect in the presence of bound states.

The point is that completeness of the eigenstates of an operator
concerns the whole spectrum, including the bound states.

In the original applications of the KLR folded diagram theory or LS
similarity transformation theory to the shell model problems, the
spectrum of the Hamiltonian is discrete and the model space is finite
dimensional. Everything looks trivial about the completeness.
In the scattering problem, however, even though one is interested in
continuous states, bound states must be included if one talks about
completeness.

One might claim that the reduced theory is designed to reproduce the
scattering amplitudes, so that it does not need to include the bound
states in the spectrum. It is however not correct because the scattering
amplitudes in general reflect some information about the bound states. A
good example is provided by the well known Levinson's
theorem\cite{LevinsonTheorem}, which states that the low-energy
scattering data (the phase shift at zero momentum) for a well-behaved
potential knows the number of the bound states. If the reduced potential
$\Vlk$ reproduces the half-on-shell, hence the on-shell T-matrix, it
must support the bound states because the existence of which is encoded
in the scattering phase shift.

If one takes into account the correct completeness relation,
Eq.~(\ref{correctcomplete}), one finds the modified RGE given by
\begin{eqnarray}
 \frac{d}{d\Lambda}\Vlk^\Lambda(k',k)
  &=&\frac{2}{\pi}
  \frac{\Vlk^\Lambda(k',\Lambda)T^\Lambda(\Lambda, k; \Lambda^2)}
  {1-(k/\Lambda)^2}
  \nonumber \\
 &&{}+\sum_i
  \int_l
  \left(
   \frac{d}{d\Lambda}\Vlk^\Lambda(k',l)
   -\frac{2}{\pi}\frac{\Vlk^\Lambda(k',\Lambda)\Vlk^\Lambda(\Lambda, l)}
   {1+\left(k_{B_i}/\Lambda\right)^2}
  \right)
  \chi^\Lambda_{B_i}(l)\left(\tilde{\chi}^\Lambda_{B_i}(k)\right)^*,
  \nonumber \\
 \label{eq:vlowkmodified}
\end{eqnarray}
where $\chi^\Lambda_{B_i}(k)=\bra k|\chi^\Lambda_{B_i}\ket$ is the wave
function in momentum space for the bound state, and
$\tilde{\chi}^\Lambda_{B_i}(k)$ is its bi-orthogonal
conjugate. 

Interestingly, however, the extra terms are shown to cancel each
other~\footnote{The demonstration below is essentially due to Bogner,
Furnstahl, and Schwenk~\cite{BFS}.}, if we assume that the state
$|\chi^\Lambda_{B_i} \ket$ is related to the state vector $|\Psi_{B_i}
\ket$ in the original theory,
\begin{equation}
 (H_0+V_{NN})|\Psi_{B_i}\ket =-k_{B_i}^2 |\Psi_{B_i}\ket,
\end{equation}
 through the projection,
\begin{equation}
 |\chi^\Lambda_{B_i} \ket = P |\Psi_{B_i} \ket,
  \label{reducedBS}
\end{equation}
in conjunction with Eq.~(\ref{stateprojection}). Note that the
conditions Eqs.~(\ref{reducedBS}) and (\ref{stateprojection}) are
independent.

To see the cancellation, let us first note that the condition
\begin{equation}
 \frac{d}{d\Lambda}\bra k' | \Vlk^\Lambda |\chi^\Lambda_{B_i}\ket=0,
  \quad (0\le k' < \Lambda),
  \label{cutoffindepBS}
\end{equation}
and Eq.~(\ref{reducedBS}) are equivalent, as Eqs.~(\ref{hosTinv}) and
(\ref{stateprojection}) are.  Then, from the bound-state
Lippmann-Schwinger equation,
\begin{equation}
 |\chi^\Lambda_{B_i}\ket=\int_p |p\ket \frac{-1}{p^2+k_{B_i}^2}
  \bra p | \Vlk^\Lambda |\chi^\Lambda_{B_i}\ket,
\end{equation}
 we have
\begin{equation}
 \frac{d}{d\Lambda}|\chi^\Lambda_{B_i}\ket 
  =-\frac{2}{\pi} |\Lambda\ket \frac{\Lambda^2}{\Lambda^2+k_{B_i}^2}
  \bra \Lambda | \Vlk^\Lambda |\chi^\Lambda_{B_i}\ket,
\end{equation}
which corresponds to Eq.~(\ref{derstate}).
From these equations, we have
\begin{eqnarray}
 0
 &=&\bra k' |\frac{d\Vlk^\Lambda}{d\Lambda}|\chi^\Lambda_{B_i}\ket
  -\frac{2}{\pi}
  \frac{
  \bra k'|\Vlk^\Lambda\ket
  \bra \Lambda |\Vlk^\Lambda|\chi^\Lambda_{B_i}\ket
  }{
  1+\left(k_{B_i}/\Lambda\right)^2
  }
  \nonumber \\
 &=&\frac{2}{\pi}
  \int_l
  \left(
   \frac{d}{d\Lambda}\Vlk^\Lambda(k',l)
   -\frac{2}{\pi}\frac{\Vlk^\Lambda(k',\Lambda)\Vlk^\Lambda(\Lambda, l)}
   {1+\left(k_{B_i}/\Lambda\right)^2}
  \right)\chi^\Lambda_{B_i}(l),
\end{eqnarray}
thus, the extra terms in Eq.~(\ref{eq:vlowkmodified}) vanish
identically.

In conclusion, though the first derivation in Ref.~\cite{Bogner:2001jn}
does not take into account the existence of bound states at all, and the
completeness relation is wrong when there are bound states, the
RGE~(\ref{eq:vlowk}) is shown to be unmodified irrespective of the
existence of bound states, if we assume  Eq.~(\ref{reducedBS}).

\section{Spectral representation for the $T$ matrix}
\label{sec:spectral}

The fourth derivation of the RGE is given in the appendix of
Ref.~\cite{Bogner:2001jn}. The derivation is based on the following
spectral representation for the T-matrix,
\begin{equation}
 \Vlk^\Lambda(k',k)=T(k',k;k^2)+ \int_p T(k',p;p^2)\frac{1}{p^2-k^2}
  T^\Lambda(p,k;p^2).
  \label{spectral-T}
\end{equation}


In the following, I will show how this representation is modified in the
presence of bound states.  Let us start with Eq.~(\ref{T-Vlk}) in a bit
different representation,
\begin{equation}
 T(k',k;k^2)=\Vlk^\Lambda(k',k) + \bra k' | \Vlk^\Lambda G_0^{(P)}(k^2)
  T(k^2) | k\ket,
  \label{T-Vlk-2}
\end{equation}
where $G_0^{(P)}(k^2)$ is the free Green function in the $P$-space,
\begin{equation}
 G_0^{(P)}(k^2)=\frac{P}{k^2-H_0}.
\end{equation}
Eq.~(\ref{T-Vlk-2}) may be rewritten by using the full Green function in
the $P$-space,
\begin{equation}
 G^{(P)}(k^2)=\frac{P}{k^2-H^\Lambda}, \quad H^\Lambda=H_0+\Vlk^\Lambda,
\end{equation}
as
\begin{equation}
 T(k',k;k^2)=\Vlk^\Lambda(k',k) + \bra k' | \Vlk^\Lambda G^{(P)}(k^2)
  \Vlk^\Lambda | k\ket.
\end{equation}
If the operator $J$ in Eq.~(\ref{J}) were the identity operator in the
$P$-space, then, inserting $J$ and using $H^\Lambda |\chi^\Lambda_p \ket
= p^2 |\chi^\Lambda_p \ket$, one would get
\begin{equation}
 T(k',k;k^2)=\Vlk^\Lambda(k',k)
  +\int_p \bra k' |\Vlk^\Lambda|\chi_p^\Lambda\ket
  \frac{1}{k^2-p^2}\bra \tilde{\chi}_p^\Lambda |\Vlk^\Lambda |k\ket,
\end{equation}
hence Eq.~(\ref{spectral-T}), with the identification $\bra k'
|\Vlk^\Lambda |\chi_p^\Lambda \ket = T(k',p;p^2)$ and $\bra
\tilde{\chi}_p^\Lambda | \Vlk^\Lambda | k \ket = T^\Lambda(p,k;
p^2)$. As we have shown, however, the operator $J$ is actually not the
identity operator in the $P$-space in the presence of bound states, this
manipulation is not justified. In the presence of bound states, the
correct equation is
\begin{eqnarray}
 T(k',k;k^2)&=&
  \Vlk^\Lambda(k',k)
  +\int_p T(k',p;p^2) \frac{1}{k^2-p^2}T^\Lambda(p,k;p^2)
  \nonumber \\
 &&+\sum_i\bra k'|\Vlk^\Lambda|\chi^\Lambda_{B_i}\ket
  \frac{1}{k^2+k_{B_i}^2}
  \bra \tilde{\chi}^\Lambda_{B_i}|\Vlk^\Lambda|k\ket.
\end{eqnarray}
Similarly, the equation which relate the left-on-shell T-matrix to the
right-on-shell one is also modified as
\begin{eqnarray}
 T^\Lambda(p,k;p^2)&=&T(p,k;k^2)
  +\int_q T(p,q;q^2)
  \left\{
   \frac{1}{p^2-q^2}-\frac{1}{k^2-q^2}
  \right\}
  T^\Lambda(q,k;q^2)
  \nonumber \\
 &&{}+\sum_i
  \bra p |\Vlk^\Lambda|\chi^\Lambda_{B_i}\ket
  \left\{
   \frac{1}{p^2+k_{B_i}^2}-\frac{1}{k^2+k_{B_i}^2}
  \right\}
  \bra \tilde{\chi}^\Lambda_{B_i}|\Vlk^\Lambda|k\ket.
\end{eqnarray}

The extra terms are however cutoff independent because of
Eq.~(\ref{cutoffindepBS}). Thus they do not contribute to the RGE for
the $\Vlk$.

\section{Summary and discussions}
\label{sec:summary}

In this paper, we have shown that each of the existent derivations of
the RGE, Eq.~(\ref{eq:vlowk}), has a serious defect. The derivations
based on the ``semi-group composition law'' are shown to be unjustified,
while the derivations based on the completeness must be modified when
bound states are present but the resulting RGE is shown to be
unmodified. After all, the present work shows that, even though the
derivation given in Ref.~\cite{Bogner:2001jn} does not take into account
the effects of bound states, Eq.~(\ref{eq:vlowk}) is correct
irrespective of the existence of bound states, if we assume
Eq.~(\ref{reducedBS}) or, equivalently, Eq.~(\ref{cutoffindepBS}).

It is important to note that, in order to derive Eq.(\ref{eq:vlowk}),
one actually needs to require not only that the half-on-shell T-matrix
is preserved, but also that the matrix elements $\bra p | \Vlk^\Lambda
|\chi^\Lambda_{B_i}\ket$ ($0\le p < \Lambda$) is invariant under the
change of the cutoff, $\Lambda$. This latter requirement does not
immediately follow the former. In this sense, the derivation rests on
the stronger (and consistent) requirements, Eqs.~(\ref{stateprojection})
and (\ref{reducedBS}), than just requiring the preservation of the
half-on-shell T-matrix.

It may be interesting to note that the RGE is not directly derived from
the concrete reduction methods such as LS and KLR methods, because the
second and the third derivations of Ref.~\cite{Bogner:2001jn} are now
shown to be invalid. Still, because LS method preserves the
half-on-shell T-matrix (and the matrix elements for the bound states),
it should satisfy the same RGE.

It is also interesting to note that the RGE for $V_{WRG}$ is much
simpler and more easily derived than that for $\Vlk$.  Furthermore,
since the $\Vlk$ RGE is too complicated to get a simple picture about
the action under the RG transformations, the usual concepts such as
relevant and irrelevant operators do not seem to be useful, while they
are actually important in the Wilsonian RG approach~\cite{Birse:1998dk,
Harada:2005tw, Harada:2006cw, Harada:2007ua}. To our best knowledge, no
one has ever found the nontrivial fixed point of the RGE for $\Vlk$,
which plays an essential role for $V_{WRG}$. We thus suspect that the
convergence of ``realistic potentials'' to a universal one is not a
(direct) consequence of ``universality'' of the RG action.

We conjecture that the convergence is best understood in the light of
the inverse scattering problem. (See Chapter 20 of
Ref.~\cite{Newton:1982} or Chapter 12 of Ref.~\cite{AlfaroRegge:1965}
for reviews.) Note that all of the ``realistic'' potentials contain the
information about the phase shifts below $E_{lab}\alt 350$ MeV, and that
the LS reduction preserves the half-on-shell T-matrix elements (hence
the phase shifts), as we will show below. Since the reduced theory does
not contain high energy scattering by construction, the information
about the phase shifts contained in the reduced potential provides more
complete information about the phase shifts ``at all energies'' as the
cutoff is lowered. On the other hand, a result of the inverse problem,
in its simplest case, is that the potential is uniquely determined by
the phase shifts at all energies. Even though careful analysis is needed
to establish the connection between the convergence and the uniqueness
of the potential in the inverse problem, it seems to provide a more
natural way of understanding the convergence than the RG
``universality.''

In Ref.~\cite{Bogner:2006vp} it is reported that the conventional way of
calculating $\Vlk$ shows a slow convergence and a new method based on
$V_{WRG}$ is proposed. It seems, however, difficult to transform
$V_{WRG}$ to an energy-independent potential analytically, even though
there may be no difficulty in doing so numerically. In
Ref.~\cite{Harada:2005tw}, such a transformation was considered for a
very simple potential as a field redefinition in the path integral
formulation. We found that there is a nontrivial Jacobian, which cannot
be calculated easily.

Recently, there emerges a new approach~\cite{Bogner:2006pc,
Bogner:2007jb, Jurgenson:2007td, Anderson:2008mu} based on the
Wegner-Glazek-Wilson (WGW) similarity RG
transformation~\cite{Wegner:1994, Glazek:1993rc, Glazek:1994qc}. In this
formulation, the RGE is given from the outset so that there is no
problem with the derivation of the RGE. It is however unclear to us if
the WGW similarity transformation preserves the half-on-shell
T-matrix. Note that the LS similarity transformation preserves the
half-on-shell T-matrix because of the particular property of the ``wave
operator'' $\omega$, $P\omega =0$, which leads to
\begin{equation}
 V_{LS}\equiv 
  P(1-\omega)(H_0+V_{NN})(1+\omega)P -PH_0P =PV_{NN}(1+\omega)P,
\end{equation}
so that
\begin{eqnarray}
 \bra k|V_{LS}|\chi^\Lambda \ket 
  &=& \bra k|P V_{NN}(1+\omega)P |\chi^\Lambda\ket
  = \bra k | V_{NN} |\Psi\ket,
\end{eqnarray}
since $Q|\Psi\ket = Q\omega P |\Psi \ket =Q \omega P|\chi^\Lambda\ket$.
On the other hand, WGW similarity transformed potential $V_s$ is defined
as
\begin{equation}
 V_{s}\equiv U(s)(T_{\scriptsize\mbox{rel}}+V)U(s)^\dagger 
  -T_{\scriptsize\mbox{rel}},
\end{equation}
where the operator $U(s)$ is chosen in most papers as
\begin{equation}
 \frac{dU(s)}{ds}U^\dagger(s)=[T_{\scriptsize\mbox{rel}}, V_s].
\end{equation}
One would be interested in the T-matrix of the transformed theory
defined as $\bra k | V_s |\chi_s\ket $, where $|\chi_s\ket \equiv
U(s)|\Psi\ket $ is the eigenstate of the transformed Hamiltonian. Unlike
the LS transformation, the T-matrix does not seem to be invariant under
the change of $s$.

\begin{acknowledgments}
 The author would like to thank NCTS for the financial support (grant
 number: NSC96-2119-M-007-001) for his stay at Taipei, during which the
 initial stage of the present work is done. He is grateful to C.-W.~Kao
 and S.-N.~Yang for the discussions and for the kind hospitality
 extended to him. He thanks S.~X.~Nakamura for the discussions via
 e-mails, and R.~Okamoto for the discussions about the model-space
 reduction methods. Discussions with H.~Kubo are also acknowledged.
 Finally, the author would like to thank S.~Bogner, D.~Furnstahl, and
 A.~Schwenk for pointing out the cancellation of the extra terms.
\end{acknowledgments}

\bibliography{NPRG,NEFT,EffInt,POTSCAT}

\end{document}